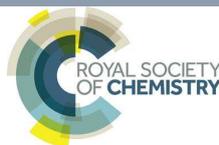

# Length segregation in mixtures of spherocylinders induced by imposed topological defects

Elshad Allahyarov[abc][†] and Hartmut Löwen[d]

We explore length segregation in binary mixtures of spherocylinders of lengths $L_1$ and $L_2$ with the same diameter $D$ which are tangentially confined on a spherical surface of radius $R$. The orientation of spherocylinders is constrained along an externally imposed direction field on the sphere which is either along the longitude or the latitude lines of the sphere. In both situations, integer orientational defects at the poles are imposed. We show that these topological defects induce a complex segregation picture also depending on the length ratio factor $\gamma = L_2/L_1$ and the total packing fraction $\eta$ of the spherocylinders. When the binary mixture is aligned along longitudinal lines of the sphere, shorter rods tend to accumulate at the topological defects of the polar caps whereas longer rods occupy central equatorial area of the spherical surface. In the reverse case of latitude ordering, a state can emerge where longer rods are predominantly both in the cap and in the equatorial areas and shorter rods are localized in between. As a reference situation, we consider a defect-free situation in the flat plane and do not find any length segregation there at similar $\gamma$ and $\eta$, hence the segregation is purely induced by the imposed topological defects. It is also revealed that the shorter rods at $\gamma=4$ and $\eta \geq 0.5$ act as obstacles to the rotational relaxation of the longer rods when all orientational constraints are released.

## 1 Introduction

In binary soft matter systems, segregation of different sort of particles can occur upon a change of the thermodynamic or environmental conditions. The equivalent in bulk equilibrium thermodynamics is the phenomenon phase separation or demixing which implies a two-phase coexistence. Whether or not phase separation occurs in the bulk for classical particles depends largely on the interactions between the particle species as well as on temperature and the partial densities. One simple "athermal" interaction is an excluded volume (or steric) interaction between two hard bodies where temperature scales out as it only trivially sets the energy unit $k_B T$. Phase separation in binary mixtures of hard particles of different shape occurs for various combinations of shapes ( see, e.g.[1]). One important and traditional examples are hard sphere mixtures[2] which do demix for certain size asymmetries. Another example are hard rod-like particle mixtures which have been recently considered in various situations[3–9] and exhibit also bulk segregation[10–12]. These are more complex than spheres since they possess an additional orientational degree of freedom.

When particles are confined on a curved manifold, the segregation and phase separation is strongly affected by the underlying curvature of configuration space. In fact, the influence of curvature on phase separation has been explored in different contexts including the crystallization transition of spheres[13–16] and segregation in two-component vesicles[17]. Phase separation was studied in curved bilayer membranes[18,19]. A systematic analysis on the impact of curvature on phase separation was performed by computer simulations of the Widom-Rowlinson model[20] and a theoretical Ginzburg-Landau approach on the sphere[21].

[a] Theoretische Chemie, Universität Duisburg-Essen, D-45141 Essen, Germany. E-mail: elshad.allahyarov@case.edu
[b] Theoretical Department, Joint Institute for High Temperatures, Russian Academy of Sciences (IVTAN), 13/19 Izhorskaya street, Moscow 125412, Russia.
[c] Department of Physics, Case Western Reserve University, Cleveland, Ohio 44106-7202, United States.
[d] Institut für Theoretische Physik II: Weiche Materie, Heinrich-Heine Universität Düsseldorf, Universitätsstrasse 1, 40225 Düsseldorf, Germany
† Present address: Theoretische Chemie, Duisburg-Essen University, Universitätsstrasse 5, 45141 Essen, Germany.

For rod-like particles tangentially confined to a sphere there is not only a pure curvature effect but there are more complex options by constraining the orientations along an imposed director field. Due to the compact topology of the sphere, a tangential director field is never defect-free but has to exhibit topological defects of the orientation[22–28]. The two simplest cases arise if the orientation field is prescribed either along the longitudinal or along the latitudinal directions of the sphere, see Figure 1. Then, in both cases, two integer topological defects arise at the two poles. An interesting question concerns the impact of topological defects in the constrained director field on length segregation. This is important both from a fundamental point of view since it links topology and thermodynamics and for actual applications as it enables to tailor segregated states at wish[29,30] by imprinting an orientational field externally[31].

In this paper we explore the impact of imposed topological defects on segregation in binary mixtures of hard rods of different lengths. We confine the particles tangentially on a spherical surface and align their orientation along certain prescribed directions which possess two integer-defects at the poles. The one-component case was studied previously both at high packings[32] and intermediate densities[33]. Here we show indeed that the presence of defects can induce length segregation at particular values of the length ratio factor $\gamma = L_2/L_1$ and the total packing fraction of the spherocylinders $\eta$. When the alignment is along the longitudinal lines of sphere, see Figure 1(a), shorter rods in the mixture accumulate at the polar caps of sphere around the defects whereas longer rods occupy the equatorial area of the spherical surface. In the second case, when the alignment is along latitudinal lines, see Figure 1(b), the segregation behavior becomes even more rich involving in particular a state where longer rods predominantly are both in the caps around the defects and in the equatorial area, and shorter rods are localized between these two areas. Conversely, in a reference situation of a flat plane without any defects, there is no length segregation at similar $\gamma$ and $\eta$, proving that the segregation is purely induced by the imposed topological defects.

We also reveal that the shorter rods act as obstacles to the rotational relaxation of longer rods at $\gamma=4$ and $\eta \geq 0.5$. Such behavior ensues the preservation of the initially ordered configuration of the longer rods when the orientational constraint is released. The jamming of longer rods, which cannot be resolved in the time window of current simulations, stipulates the implementation of additional methods for overcoming energy barriers to reach true equilibrium.

The remainder of this paper is structured as follows. The details of our simulation method for the spherocylinders anchored on a spherical surface are given in section 2. In section 3 we discuss simulation results for the length segregation in binary mixtures of spherocylinders with prescribed longitude ordering. The segregation process in binary mixtures aligned along latitude lines is analyzed in section 4. Section 5 is devoted to the details of the ordering preservation of longer rods in relaxed binary mixtures. We conclude in section 6.

## 2 Simulation Model

An equimolar binary mixture of hard-core spherocylinders consisting of $N/2$ rods of length $L_1$ and $N/2$ rods of length $L_2$ is anchored on the spherical surface $S_2$ of diameter $2R$. All rods have the same diameter $D$. The end-to-end length of the spherocylinder is $L_i + D$ ($i=1,2$), where $D$ accounts for the hemispherical caps at the ends of the rod, see Figure 1(d). The rods interact through a hard-core potential

$$u_{ij}(\vec{R}_i, \vec{R}_j, \vec{n}_i, \vec{n}_j) = \begin{cases} \infty & \text{if } i \text{ and } j \text{ overlap}, \\ 0 & \text{otherwise}. \end{cases} \quad (1)$$

where $\vec{R}_i$ is the anchoring positions of rod $i$, see Figure 1(e).

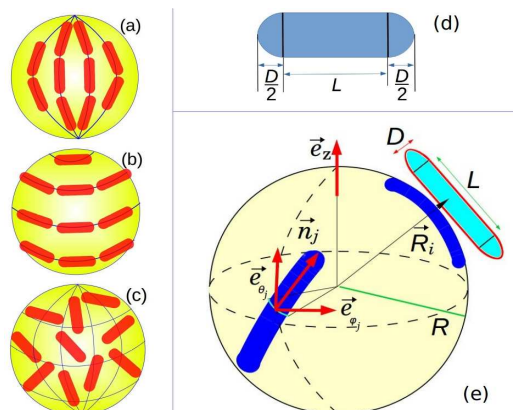

**Fig. 1** Schematic pictures explaining simulation model. (a)- a system with prescribed longitudinal orientation, (b)- a system with prescribed latitudinal orientation, (c)- a relaxed freely rotated system, (d)- a spherocylinder of length $L$ and diameter $D$, (e)- a projection of spherocylinder with index $i$ on the spherical surface and orientation vector $\vec{n}_j$ of another rod $j$. The unit vectors $\vec{e}_z$, $\vec{e}_{\theta_j}$, and $\vec{e}_{\varphi_j}$ correspond to the unit vector along the the $z$-axis, and the unit polar and azimuthal angle vectors of rod $j$, respectively

A solid angle $\hat{\Omega}_i = \hat{\Omega}(\theta_i, \varphi_i)$ describes the orientational part of the particle position $\vec{R}_i$ on the sphere and is given by the spherical azimuthal angle $\varphi_i$ and the polar angle $\theta_i$. The anchoring position for the $i$-th particle on $S_2$ is given by its radius-vector $\vec{R}_i = \left(R + \frac{D}{2}\right)(\sin\theta_i \cos\varphi_i, \sin\theta_i \sin\varphi_i, \cos\theta_i)$, which points from the center of the host sphere to the geometrical center of the spherocylinder. The anchoring is imposed by maintaining $|\vec{R}_i| = R + D/2$ at all times. The orientation of the spherocylinder at the position $\vec{R}_i$ is given by a unit vector $\vec{n}_i$, which is directed along the long axis of the spherocylinder and is perpendicular to $\vec{R}_i$, see Figure 1(e).

The areal packing fraction of the spherocylinders on $S_2$ is defined as $\eta = \frac{1}{2}N\left(S_p(1) + S_p(2)\right)\left(\pi(2R+D)^2\right)$, where $S_p(i)$ is the shadow area of the spherocylinder $i$ on $S_2$, see for details Ref.[33]. In the following we fix the radius of the host sphere to $R=70D$.

Langevin dynamics simulations were carried out for different binary mixtures of spherocylinders with packing fractions $\eta$ changed between 0.3 and 0.85, the length rate factor $\gamma = L_2/L_1$ varied between 1 and 4, an the rod aspect ratio $L_i/D$ considered between 4 and 24. These parameters correspond to the total number of particles $N$ on the $S_2$ taking values between $10^3$ for low density simulations and $10^4$ for high density simulations.

During the simulation runs each Langevin move consisted of positional and orientational displacement steps for particles. The first step changes the rod position from its old anchoring point $\vec{R}_i$ to the new position $\vec{R}_i + \delta\vec{R}_i$. The second step changes the rod orientation $\vec{n}_i$ to the new orientation $\vec{n}_i + \delta\vec{n}_i$ through the rotation of the rod around $\vec{R}_i$. During all Langevin moves the orientation $\vec{n}_i$ is always perpendicular to $\vec{R}_i$, $\vec{n}_i \cdot \vec{R}_i = 0$. Both the displacements $\delta\vec{R}_i$ and $\delta\vec{n}_i$ are taken from relevant Gaussian distributions. The full details of the Langevin moves are described in Ref.[33].

We consider two basic preordered configurations which we refer to as the *longitude* and *latitude* oriented orderings, for the spherocylinders on the sphere, see Figure 1(a) and 1(b). For the preordered longitude ordering the rod orientation vector $\vec{n}_i$ obeys $\vec{n}_i \cdot \vec{e}_\theta = 1$, where the polar angle vector is defined as $\vec{e}_\theta = \vec{R}_i \times \left[\vec{e}_z \times \vec{R}_i\right]$. In a similar way, for the preordered latitude ordering the rod orientation is fixed in accordance with $\vec{n}_i \cdot \vec{e}_\varphi = 1$, where the azimuthal angle vector is $\vec{e}_\varphi = [\vec{e}_z \times \vec{R}_i]$. Here $\vec{e}_z$ is a unit vector along the $z$-axis.

After each attempted Langevin move the new position and orientation for the $i$-th rod are accepted if the minimal distance between the rod and its nearest neighbor particle $j$, calculated using the procedure described in Ref.[34], obeys $r_{ij}^{(m)} \geq D$. In the opposite case, the particle is kept at its old position and orientation. For the better treating of the particle-particle overlappings we use $h=10^{-3}\tau$ as a time step in low density simulations and $h=10^{-4}\tau$ in high density simulations. The characteristic time $\tau$ for the spherocylinder center of mass is defined as a time needed to cover a distance $D$, $\tau \approx D\sqrt{m/(k_B T)}$. Though a more refined procedure[35] can be implemented for defining the exact impact time between the particles and calculating their expected new positions and orientations after each collision step, this procedure greatly slows down the simulation without any palpable gain in the precision of simulations.

Each simulation run was started with random insertions of rods on $S_2$ under the imposed angular constraint. While the $i$-th particle is inserted, all other $i$-1 particles were allowed to undergo Langevin moves on $S_2$, in the spirit of of Ref.[33]. This procedure increases the effectiveness of the particle insertion at high packing fractions $\eta$. Once the desired $\eta$ is achieved, the system is equilibrated during $10^7$-$10^8$ Langevin steps corresponding to $10^4\tau$–$10^5\tau$ simulation times. In the following the same amount of simulation steps were run to gather necessary statistics for the production phase during which the density $\rho(\vec{r})$ of the shorter and longer particles on the $S_2$ was calculated.

Each of the completed simulations for preordered mixtures were subsequently relaxed by lifting the imposed angular restriction and giving the particles the ability to freely rotate around their radius-vectors $\vec{R}_i$, see Figure 1(c). These free ordered Langevin simulations were run until a new virtually equilibrated state is reached (also referred hereafter as a freely rotated state). The purpose of these freely rotated simulations was to prove that the length segregation is only possible in the preordered state and completely disappears in the freely rotated state.

In the following sections we will analyze how the presence of the defects enhances the demixing of the binary mixture into the short-rod rich and the long-rod rich zones depending on the total packing fraction of particles $\eta$ and the length ratio factor $\gamma$. In all presented simulations the mixture composition was kept at 0.5 meaning that half of the rods has shorter length $L_1$ and the other half of the rods has longer length $L_2$. It should be mentioned that the equal composition of the mixture does not guarantee the equal partial packing fractions for its components. For long rods, in the binary mixture with the length ratio factor $\gamma$, the partial packing fraction of the shorter component is about $\gamma$ times smaller than the partial packing fraction of the longer component.

## 3 Demixing in binary mixture with $\vec{n}_i \parallel \vec{e}_{\theta_i}$

We first start with the binary mixture aligned along longitudinal lines of the spherical surface $S_2$. The state diagram of the observed structures corresponding to this ordering with representative snapshots and schematic pictures are collected in Figure 2. Corresponding plots for the polar particles densities $\rho_1(\theta)$ for the shorter rods and $\rho_2(\theta)$ for the longer rods are shown in Figure 3.

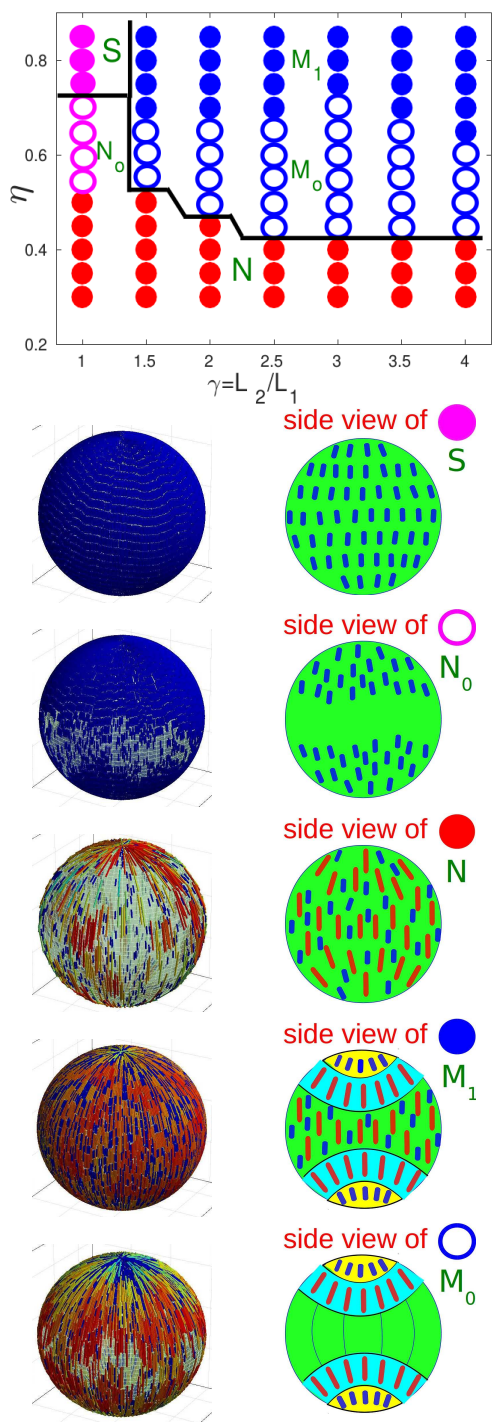

The state diagram in Figure 2 distinguishes five different states possible for the binary mixture of rods as a function of $\eta$ and $\gamma$. There is a single high-density smectic phase for the monodisperse system denoted as S, its representative snapshot and schematic picture are shown in the first line of Figure 2. In the snapshot all the rods for the monodisperse system with $\gamma=1$ are colored in blue. The existence of such smectic phase was discovered in our previous work in Ref.[33]. Because of the fact that $\gamma=1$, no length segregation exists for this state. The particle distribution $\rho_1(\theta)$ for this state is almost homogeneous, see the black line with circles in the left column and third row picture of Figure 3.

Fig. 2 State diagram for the binary mixture of rods with prescribed longitude ordering. Black lines separate the smectic S and nematic $N_0$ and N states from the segregation structures $M_0$ and $M_1$. Simulation snapshots and schematic pictures for the five representative structures are given below the diagram. In the snapshots shorter rods are colored blue. The coloring of longer rods, from light blue to dark red, corresponds to the strength of nematic ordering in their surrounding.

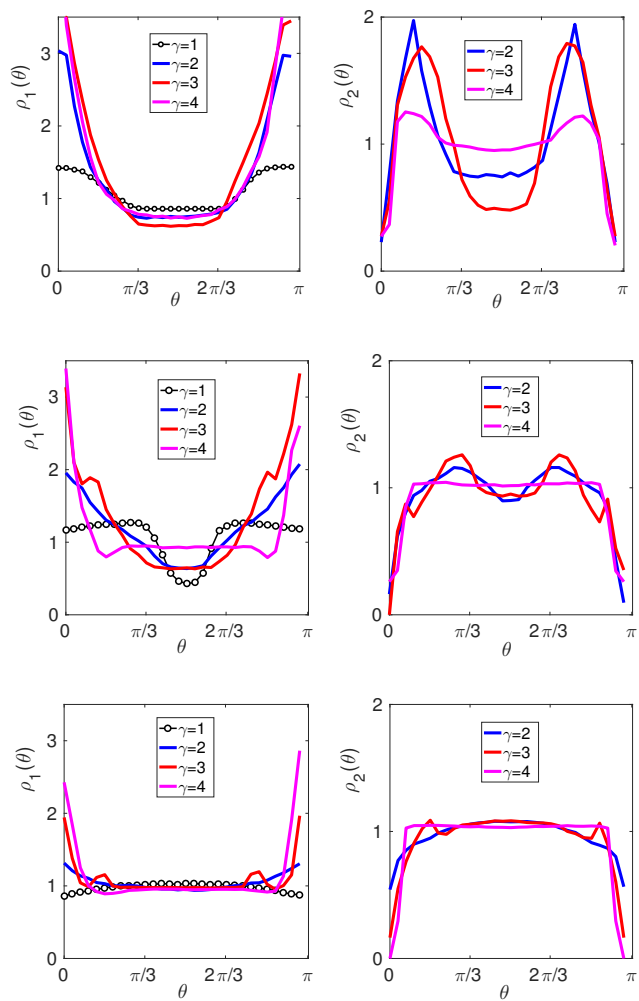

Fig. 3 Longitudinally ordered binary mixture with the shorter rod length $L_1=6D$. Left column: polar density $\rho_1(\theta)$ for the shorter rods. Right column: polar density $\rho_2(\theta)$ for the longer rods. Black line with symbols- $\gamma=1$, blue line- $\gamma=2$, red line- $\gamma=3$, pink line- $\gamma=4$. The first row- $\eta=0.5$, the second row- $\eta=0.7$, and the third row- $\eta=0.85$

Below the smectic phase at the same $\gamma=1$, there is a medium density nematic phase for the monodisperse system denoted as $N_0$. This state, also first discovered in Ref.[33], has a particle-free zone in the central area of $S_2$, see the second line in Figure 2. In the snapshot shorter rods are colored blue, while the coloring of longer rods depends on nematic ordering in their surrounding. Light blue colored longer rods have less nematic ordering than dark red colored rods. A detailed analysis shows that the polar areas of $N_0$ contain partially smectic clusters of particles. Whereas this state has no length segregation because of $\gamma=1$, there is however a nonhomogeneous particle distribution on $S_2$ as seen from Figure 3, the black line with symbols in the left column and second row picture. The particle density $\rho_1(\theta)$ reaches a minimum at $\theta=\pi/2$ meaning that the rods tend to gather at the poles leaving the equatorial area of $S_2$ sparsely occupied.

The other nematic phase in the state diagram is denoted as N and occurs at low densities for all mixtures. A representative snapshot and schematic picture for this state are provided in the third line of Figure 2. For the low values of $\gamma$ the distribution of shorter and longer particles are almost homogeneous on $S_2$, whereas at higher $\gamma$ a particle clustering, especially for longer particles with the length $L_2$, develops in the system. These clusters move freely over $S_2$ and thus, when averaged over longer simulation times, the particle densities $\rho_1(\theta)$ and $\rho_2(\theta)$ do not show any robust demixing tendencies.

The other two states in Figure 2, denoted as $M_1$ for high density mixtures and $M_0$ for low density mixtures are the newly found states with strong particle segregation abilities. Both of these states show a distinct demixing of the mixture to the shorter rod abundant zone at the poles, and the longer rod abundant horizontal zone between the pole and the central area of $S_2$. These zones are shown in the fourth and fifth lines of of Figure 2 as yellow and cyan colored areas, respectively. The formation of these zones is visible in the polar densities of particles shown in Figure 3. The left column of this figure shows a strong accumulation of the shorter rods at the poles, see the blue, red, and pink lines reaching maximum values at $\theta=0$ and $\theta=\pi$. The right column shows the polar densities of longer particles with a double maximum on the $S_2$. The strength of the segregation is high at $\eta=0.5$ for the low-density $M_0$ state compared to the high-density state $M_1$. At the same time, for $\eta<4$ the segregation disappears as the state $M_0$ is replaced by the state N.

The main difference between the $M_1$ and $M_0$ segregation states is the existence of empty patches at the central area of $S_2$ in the latter case. Also, in the $M_1$ state the densities of the shorter and longer rods match each other in the central area of $S_2$. It should be noted that the length segregation in the $M_1$ and $M_0$ states stems exclusively from the topological defects imposed in the host surface $S_2$. This is proved by the inability of a planar defect-free surface to keep the initially segregated structure of the mixture intact, see the results of our additional simulations presented in Appendix A.

## 4 Demixing in binary mixture with $\vec{n}_i \parallel \vec{e}_{\varphi_i}$

In this section we analyze the segregation process in latitude ordered binary mixture. The state diagram of the observed structures corresponding to this ordering with representative snapshots and schematic pictures are collected in Figure 4. Corresponding plots for the polar particles densities $\rho_1(\theta)$ and $\rho_2(\theta)$ are collected in Figure 5.

The pure smectic S and nematic N states in the state diagram for $\gamma=1$ are structurally similar to the S and N states discussed in previous section with the only difference that the particles are oriented now along the latitude lines of the $S_2$. The state $S_0$ for $\gamma=1$ with empty caps and a smectic phase in the central area of $S_2$ is schematically shown in the first line of Figure 4. The polar density of particles, see the black line with symbols in the left column of Figure 5, explicitly reveals the emptiness of the polar areas which are colored yellow in the schematic picture in Figure 4. The states S, $S_0$, and N were reported in our previous paper[33], and do not belong to the length segregation structures.

When the length ratio factor $\gamma$ increases, the $S_0$ and N states develop into the length segregated structure $E_0$. In this segregated phase, which is detected for the first time and shown schematically in the second line of Figure 4, the shorter rods accumulate near the edges of the empty cap area, and the longer rods occupy the central area of $S_2$. The blue, red, and pink lines in the first row pictures, and the blue and red lines on the second row pictures of Figure 5 correspond to the particle densities of this state.

At high packing fractions the $E_0$ segregation state transfers into either the very weekly segregated state $E_1$ at the low values of $\gamma$, or to the moderately segregated state $E_2$ at the moderate values of $\gamma$, or to the strongly segregated state $E_3$ at the high values of $\gamma$. All these states are newly discovered.

The weakly segregated state $E_1$, shown in the third line of Figure 4, has a noticeable length segregation as seen from the blue lines in the third row pictures of Figure 5. In this case there are slightly more shorter rods at the poles compared to the longer rods.

In the moderately segregated state $E_2$, shown in the fourth line of Figure 4, more shorter rods accumulate at the poles and more longer rods fill in the central area of

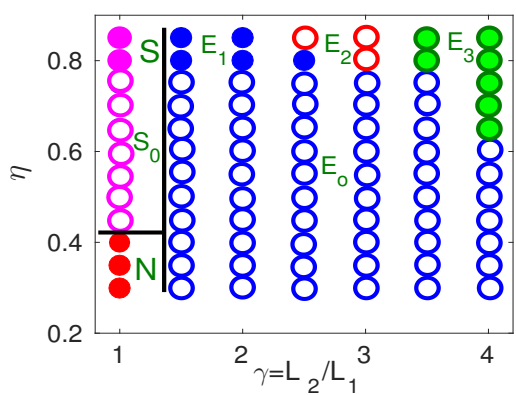

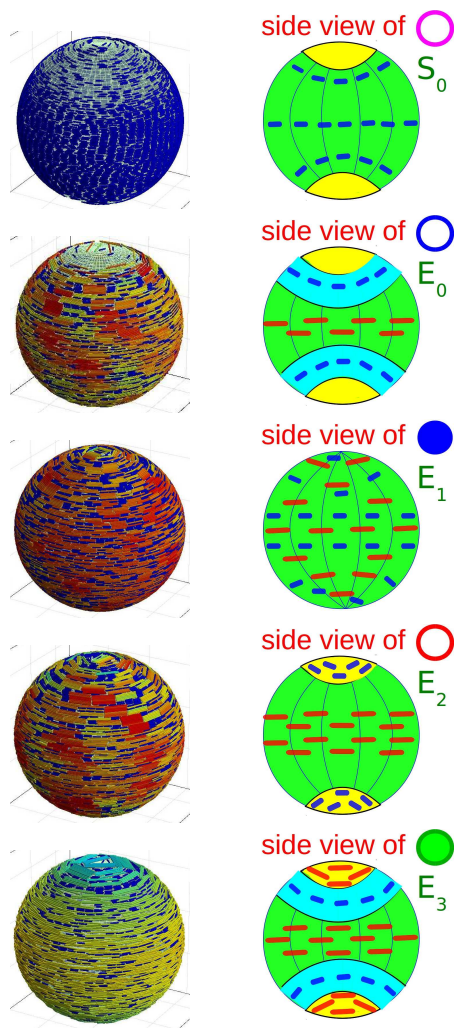

**Fig. 4** State diagram for the binary mixture of rods with prescribed latitude ordering. Black lines separate smectic S and $S_0$, and the nematic N states from the segregation structures. Simulation snapshots and schematic pictures for the five representative structures are given below the diagram.

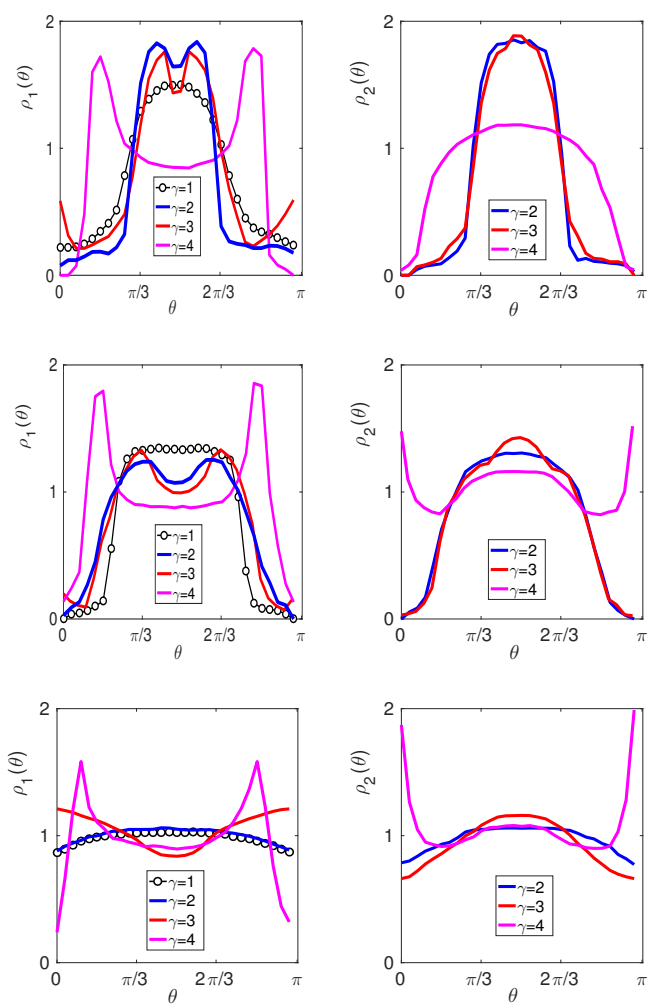

**Fig. 5** Latitudinally oriented binary mixture with the shorter rod length $L_1=6D$. Left column: polar density $\rho_1(\theta)$ for the shorter rods. Right column: polar density $\rho_2(\theta)$ for the longer rods. Black line with symbols- $\gamma=1$, blue line- $\gamma=2$, red line- $\gamma=3$, pink line- $\gamma=4$. The first row- $\eta=0.5$, the second row- $\eta=0.7$, and the third row- $\eta=0.85$

$S_2$. This is evident from the red lines in the third row picture of Figure 5.

The most intriguing state is the segregated state $E_3$ which is shown in the fifth line of Figure 4. In this state the longer rods mostly exhibit a double occupation tendency: they outnumber the shorter rods at the poles and in the central area of $S_2$. The shorter rods are sandwiched between these areas. These unique segregation picture appears only at high packing fractions $\eta \geq 0.6$ and high values of length ratio factor $\gamma \geq 3.5$.

## 5 Preserved ordering in freely rotating binary mixtures

When the angular constraints are are taken off and the rods are given the freedom to rotate around their radius vectors $\vec{R}_i$, the length segregation ability of the states are lost completely for the cases when $\gamma<4$. This can be evidenced from the snapshot pictures shown in the first column (cases (a) and (d)) in Figures 6 and 7. The upper row pictures in these figures correspond the ordered state, whereas the bottom row pictures are for the freely rotated state. In Figures 6(d) and 7(d) topological defects on the particle scale emerge but these are different from the imposed ones by prescribing the orientational ordering a priori. Furthermore, it is seen that for $\gamma=4$, the cases (b) and (c) in both figures, the freely rotated state has the longer particles preserving their initial orientation. Such behavior seems strange in the light of our previous simulation results (see Ref.[33]) showing that all the freely rotated states for the one-component ( monodisperse) rod systems with $L=24D$ to lost their initial ordering for $\eta < 0.85$. The states (b) and (c) in Figures 6 and 7, which show a freezing of the longer rods, have packing fractions $\eta=0.7$ and $0.75$, respectively. At such moderate packing fractions a monodisperse spherocylinder system with $L_1=L_2=24D$ will quickly lost its initial orientation when the prescribed orientation field is taken off, see Appendix B.

Our simulations show that the jamming behavior for $\gamma=4$ exists at even lower packing fraction $\eta \approx 0.5$ binary mixtures. We assume that the main reason for such orientation preservation in the binary mixture is the existence of shorter rods around the longer rods. The shorter rods can be viewed as obstacles which decrease the available space for the rotation of longer rods. In other words, the shorter rods increase the effective packing fraction of the longer rods which contributes to the order preservation of the latter in freely rotated configurations. On the other hand, however, it can be speculated that the observed jamming of longer rods in the binary mixture is not the true and final equilibrium state because of huge energetical barriers. Thus, additional methods should be implemented to resolve this metastable state.

## 6 Conclusions

We carried Langevin simulations for binary mixture of spherocylinders placed on the spherical surface under orientational constraints. According to our simulation data the imposed defect structure in the host surface strongly contributes to the particle segregation by their length when the mixture is aligned either along longitudinal or latitudinal lines of the sphere. The segregation depends on the length ratio factor $\gamma=L_2/L_1$ of the mixture and the total packing fraction $\eta$ of the spherocylinders. In longitudinally aligned binary mixture, shorter rods are in abundance at the polar caps of sphere whereas longer rods accumulate in the cen-tral area of the spherical surface. For the binary mixture with prescribed latitude ordering the segregation becomes more complex. We detected a process when longer rods are predominantly in the cap and central areas of sphere and shorter rods are localized between these areas. Additionally, it is revealed that the shorter rods at $\gamma=4$ and $\eta \geq 0.5$ act as an obstacle to the rotational relaxation of longer rods. Such behavior ensues the preservation of the initially ordered configuration of the longer rods. However, such preservation might disappear in much longer simulations needed to reach true and final equilibrium state for such jammed state. The time window of our current simulations seemingly are not enough for making a decisive conclusion about the fully relaxed system with $\gamma=4$ and $\eta \geq 0.5$.

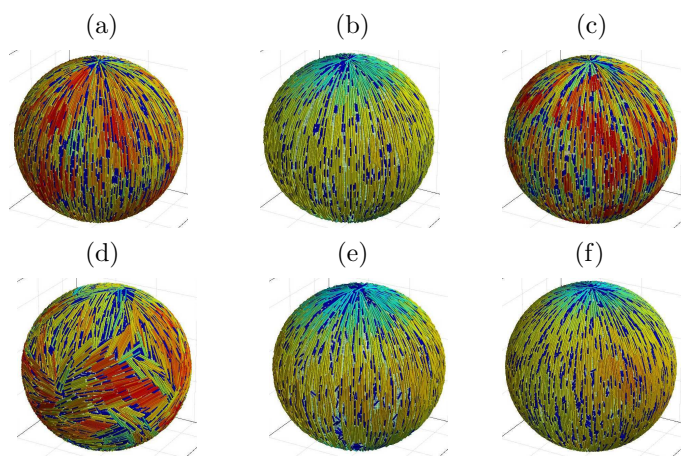

**Fig. 6** Longitudinally oriented states (a)-(c) (which correspond to the segregation structure $M_1$), and the freely rotated states (d)-(f) of the binary mixture with the following parameters: $L_1=6D$, $\gamma=3$ and $\eta=0.85$ for (a) and (d); $L_1=6D$, $\gamma=4$ and $\eta=0.7$ for (b) and (e); $L_1=4D$, $\gamma=4$ and $\eta=0.75$ for (c) and (f);

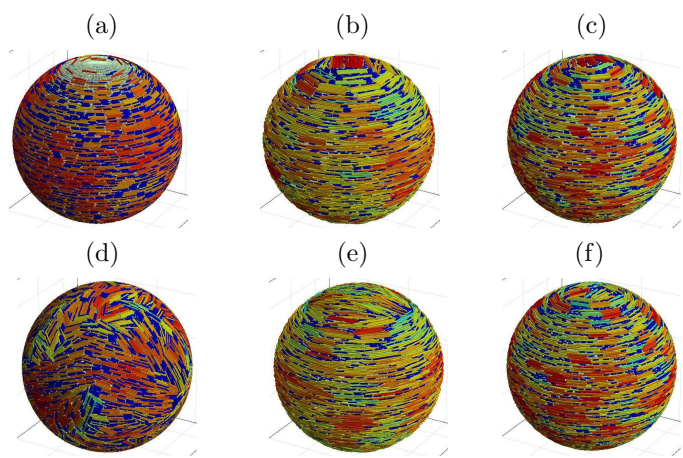

**Fig. 7** Latitudinally ordered (a)-(c), and orientationally relaxed (d)-(f) binary mixtures with the following parameters: $L_1=6D$, $\gamma=2$ and $\eta=0.7$ for (a) and (d); $L_1=6D$, $\gamma=4$ and $\eta=0.5$ for (b) and (e); $L_1=4D$, $\gamma=4$ and $\eta=0.75$ for (c) and (f);

An interesting question to answer will be how the segregation process depends on the radius of the host surface, or, more specifically, on the aspect ratios $L_1/R$ and $L_2/R$. On the one hand, for smaller radius $R$ of the $S_2$ the segregation is expected to be strong, however, in this case the density fluctuations in the mixture with a low number of particles (small surface area $S_2$ will be covered by smaller number of particles) will blur the segregation picture. On the other hand, for larger $R$, in the limit of approaching the flat surface case, no length segregation is expected as seen from Appendix A. Therefore, there should be an optimal rod-to-sphere aspect length parameter $L/R$ at which the segregation becomes more efficient. In this sense the segregation in binary mixtures resembles the nematic ordering field parameter $\sigma$ in spatially constrained nematic rods in the KKLZ model (Kutnjak, Kralj, Lahajnar, and Zumer)[36], which inversely depends on the pore radius $R$.

In conclusion, we report on the length segregation in aligned binary mixture of monodisperse spherocylinders induced by orientational topological defects imposed in the host surface. We believe that our study will attract more research toward the application of free energy based approaches to explain and predict the bigger picture on the demixing of binary mixtures under topological constraints. Fundamental measure density functional theory is one of the promising tools for such microscopic theories as it was applied to hard spherocylinders and rectangles confined on two-dimensional flat and curved manifolds even with orientational constraints[37,38]. Moreover it would be interesting how stable the segregation effect will be if the simulation model is changed in terms of interactions and set-ups[14,39–43]. Last it is worth to point out that active particles have been studied on the sphere revealing phenomena like swarm winding[44–46], aging[47] and topological sound creation[48] and length segregation is still unexplored in active systems in the presence of defects.

Finally we emphasize that the segregation behavior predicted by our simulations can in principle be verified in experiments using smectic shells of molecular crystals[49–58] or Pickering emulsion droplets covered with rod-like colloids[59–64]. Another option are layers of silica rods which are recently studied under various constraints[65,66], aspherical surfactants[67] or ellipsoidal colloids bound to curved fluid-fluid interfaces[68]. Even living and motile "particles" like cells[41] and rod-like fly embryos[69] were recently studied on spheres. An orientational constraint can be imprinted by using external fields or a molecular liquid crystal which prescribes the orientational ordering of larger colloidal rods[31].

## Acknowledgments

We thank A. Voigt for helpful discussions. We acknowledge support of this work by the Deutsche Forschungsgemeinschaft (DFG) through the grants AL 2058/1-1 (for E.A.) and LO 418/20-1 (for H.L.).

## Appendix

### A  Oriented binary mixture on a flat surface

In order to show that the length segregation in the aligned binary mixture is only possible on curved surfaces, we run three additional simulations with different length ratios on the flat surface with initially segregated mixtures. The snapshots for the fully segregated mixtures at the simulation time $t=0$ with the length ratio factor $\gamma=2$, 3, and 4 at the mixture composition 0.5 and the packing fraction $\eta=0.7$ are shown in the left column of Figure A1. The completely mixed configurations shown in the right column correspond to the simulation results at $t=10^3~\tau$. The linear densities of the shorter and longer rods along any direction show a homogeneous distribution of particles with no length segregation.

### B  Lost of ordering in a one-component system of longer rods

In order to show that the orientation order preservation for the longer rods in the binary mixture with $\gamma=4$ is indeed associated with the obstacle-like behavior of the shorter rods, we run a one-component (monodisperse) $L_1=L_2=24D$ system simulation at $\eta=0.7$ without the shorter rods. The snapshots presented in Figure B1 indicate that the monodisperse system of longer rods losses its initial ordering in the freely rotated state. A similar binary mixture with additional shorter rods $L_1=L_2/4$ would, however, retain its initial orientation.

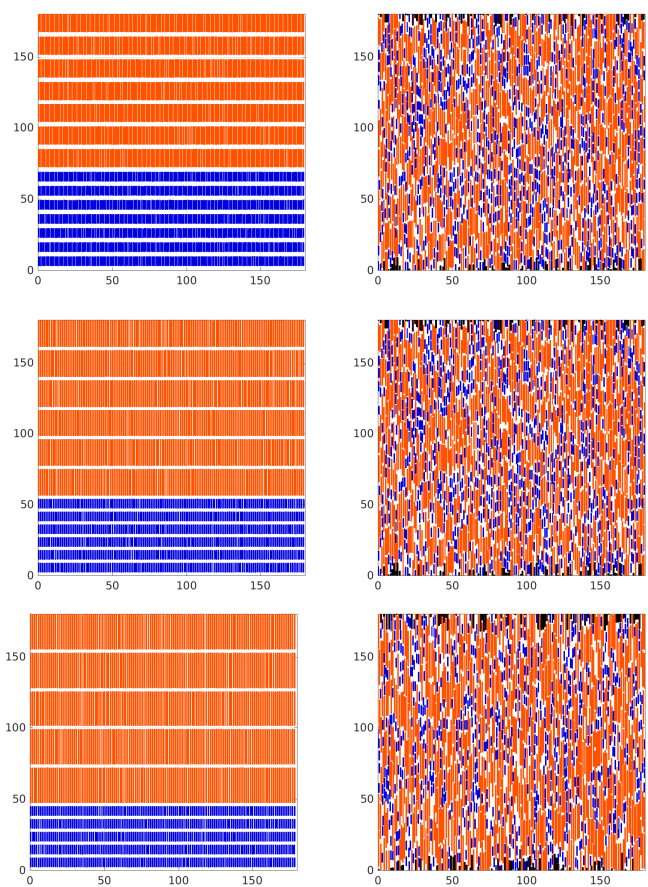

**Fig. A1** Simulations of the oriented binary mixture on a flat surface. Left column- initial configuration with sharply segregated state. Right column- equilibrated state in which the initial segregation is lost. The system parameters are: $L_1=6D$, $\eta=0.7$. First row- $\gamma=2$. Second row- $\gamma=3$. Third row- $\gamma=4$.

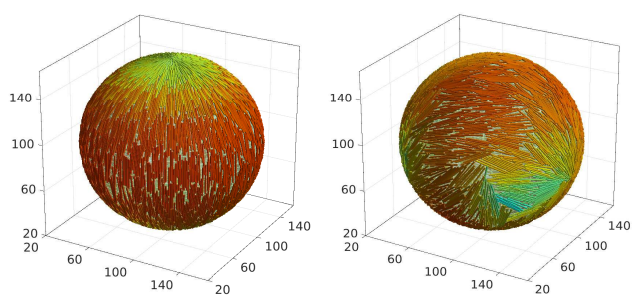

**Fig. B1** Simulation snapshots for a monodisperse system of spherocylinders with $L=24D$ and $\eta=0.7$. Left picture- a system with prescribed longitude ordering. Right picture- freely rotated system.